\documentclass{aa_new}

\usepackage{graphicx}
\usepackage{txfonts}

\newcommand{\ETAL}{{et al.}}

\newcommand{\ddd}{{\rm d}}
\newcommand{\SN}{{\rm SN}}
\newcommand{\OBS}{{\rm obs}}

\newcommand{\QUINT}{{\rm \scriptscriptstyle Q}}
\newcommand{\BCKG}{{\rm \scriptscriptstyle B}}
\newcommand{\CONST}{{\rm Const}}
\newcommand{\CRIT}{{\rm c}}
\newcommand{\PL}{{\rm \scriptscriptstyle Pl}}
\newcommand{\BAR}{{\rm b}}
\newcommand{\MAT}{{\rm mat}}
\newcommand{\MAX}{{\rm max}}
\newcommand{\TOT}{{\rm tot}}
\def\gtrsim{\;\raise 0.4ex\hbox{$>$}\kern -0.7em\lower 0.62
ex\hbox{$\sim$}\;}
\def\lesssim{\;\raise 0.4ex\hbox{$<$}\kern -0.8em\lower 0.62
ex\hbox{$\sim$}\;}
\newcommand{\UUNIT}[2]{
{\;\mathrm{#1}^{#2}} }

\newcommand{\OmM}{\Omega_\MAT}

\newcommand{\OmT}{\Omega_\TOT}
\newcommand{\Obhh}{\Omega_\BAR h^2}
\newcommand{\sig}{\sigma_8}
\newcommand{\OQ}{\Omega_\QUINT}

\begin{document}

\title{Cosmological parameters estimation \\ in the Quintessence Paradigm}

\titlerunning{Cosmological parameters estimation \\
in the Quintessence Paradigm}
\author{
M.~Douspis \inst{1, \, 2}\fnmsep\thanks{e-mail: {\tt
douspis@astro.ox.ac.uk}} \and A.~Riazuelo
\inst{3}\fnmsep\thanks{e-mail: {\tt riazuelo@spht.saclay.cea.fr}}
\and Y.~Zolnierowski \inst{1, \, 4, \, 5}\fnmsep\thanks{e-mail: {\tt
zolniero@ast.obs-mip.fr}} \and A.~Blanchard
\inst{1}\fnmsep\thanks{e-mail: {\tt alain.blanchard@ast.obs-mip.fr}}}

\institute{
Laboratoire d'Astrophysique de l'Observatoire Midi-Pyr{\'e}n{\'e}es,
14 Avenue E.~Belin, F--31400 Toulouse (France)
\and
Nuclear and Astrophysics Laboratory, Keble Road, Oxford,  OX1 3RH, (U. K.)
\and
Service de Physique Th\'eorique, CEA/DSM/SPhT, Unit\'e de recherche
associ\'ee au C.N.R.S., CEA/Saclay, F--91191 Gif-sur-Yvette c\'edex
(France)
\and
L.A.P.P., IN2P3-C.N.R.S., B.P. 110, F--74941 Annecy-le-Vieux c\'edex
(France)
\and 
Universit\'e de Savoie, B.P. 1104, F--73011 Chamb\'ery c\'edex
(France) }

\offprints{douspis@astro.ox.ac.uk}

\date{December 4 2002, {\bf Version 12}}

\date{Received ??? / Accepted ???}

\abstract{ We present cosmological parameter constraints on flat
cosmologies dominated by dark energy using various cosmological data
including the recent Archeops angular power spectrum measurements. A
likelihood analysis of the existing Cosmic Microwave Background data
shows that the presence of dark energy is not requested, in the
absence of further prior.  This comes from the fact that there exist
degeneracies among the various cosmological parameters constrained by
the Cosmic Microwave Background.  We found that there is a degeneracy
in a combination of the Hubble parameter $H_0$ and of the dark energy
equation of state parameter $w_\QUINT$, but that $w_\QUINT$ is not
correlated with the primordial index $n$ of scalar fluctuations and
the baryon content $\Omega_\BAR h^2$. Preferred primordial index is $n
= 0.95 \pm 0.05 (68\%)$ and baryon content $\Omega_\BAR h^2 = 0.021
\pm 0.003$.  Adding constraint on the amplitude of matter fluctuations
on small scales, $\sig$, obtained from clusters abundance or weak
lensing data may allow to break the degenaracies, although present-day
systematics uncertainties do not allow firm conclusions yet. The
further addition of the Hubble Space Telescope measurements of the
local distance scale and of the high redshift supernovae data allows
to obtain tight constraints. When these constraints are combined
together we find that the amount of dark energy is
$0.7^{+0.10}_{-0.07}$ ($95\%$ C.L.) and that its equation of state is
very close to those of the vacuum: $w_\QUINT < -0.75 $ ($> 95\%$
C.L.).  In no case do we find that quintessence is prefered over the
classical cosmological constant, although robust data on $\sig$ might
rapidly bring light on this important issue.  
} \maketitle

\section{Introduction}

The determination of cosmological parameters has always been a central
question in cosmology. In this respect the measurements of the
Cosmological Microwave Background (CMB) anisotropies on degree angular
scales has brought one of the most spectacular results in the field:
the flatness of the spatial geometry of the universe, implying that
its density is close to the critical density.  Although, during the
last twenty years the evidence for the existence of non-baryonic dark
matter has strongly gained in robustness, observations clearly favor a
relatively low matter content somewhere between $20$ and $50\%$ of the
critical density, indicating that the dominant form of the density of
the universe is an unclustered form. Furthermore, the observations of
distant supernovae, at cosmological distance, provide a direct
evidence for an accelerating universe, which is naturally explained by
the gravitational domination of a component with a relatively large
negative pressure, $P_\QUINT = w_\QUINT \rho_\QUINT$ with $w_\QUINT <
- 1 / 3$.  The cosmological constant $\Lambda$ (for which $w_\Lambda =
- 1$) is historically the first possibility which has been introduced
and which satisfies this requirement.  However, the presence of a
non-zero cosmological constant is a huge problem in physics: (i)
quantum field theory predicts that $\Lambda$ should be the sum of a
number of enormous contributions, so in order to avoid a cosmological
catastrophe, it is usually assumed that a yet unknown mechanism
produces a cancelation between all these contributions; (ii) it is
difficult to think of a mechanism which puts $\Lambda$ to $0$ exactly,
but it is even more difficult to find a mechanism which gives
$\rho_\Lambda \sim \rho_\CRIT \sim 10^{- 122} \rho_\PL$, as the
supernovae observations suggest, where $\rho_\PL$ and $\rho_\CRIT$ are
the Planck energy density and the critical density today,
respectively. For this reason the concept of quintessence, a scalar
field with negative pressure, has recently been proposed as a possible
alternative to a cosmological constant.
 
In this paper we shortly describe the quintessence paradigm and its
effect on some observable quantities. We then summarise the different
sets of data and methods used to constrain cosmological parameters. We
then conclude by showing the results on quintessence and cosmological
parameters.

\section{Quintessence}

The idea of quintessence was proposed in order to allow the presence
of an non zero dark energy, as suggested by observations, without
being confronted to the dramatic fine-tuning problem of the
cosmological constant.  Indeed, one still assumes that an unknown
mechanism puts the bare cosmological constant is zero, and the
smallness of the dark energy has a dynamical origin, coming from a
scalar field $\phi$ which has not yet reached the minimum of its
potential $V (\phi)$.

A large number of quintessence models were already considered in the
literature. Historically, the first proposed quintessence model had an
inverse power law potential, $V (\phi) = M^{4 + \alpha} / \phi^\alpha$
(Ratra \& Peebles \cite{ratra}; Wetterich \cite{wett}; Caldwell
\ETAL{} \cite{caldwell}), where the exponent $\alpha$ is positive and
$M$ is an energy scale fixed so that the scalar field has the correct
energy density today.  This unusual shape, which can have some
motivations from particle physics (Bin\'etruy \cite{bin1},
\cite{bin2}), insures that regardless of the initial conditions, the
field will reach a so-called ``tracking regime'' (Steinhardt \ETAL{}
\cite{steinh}), where its pressure and energy density tend to a
constant ratio given by $P_\QUINT / \rho_\QUINT \equiv w_\QUINT =
(\alpha w_\BCKG - 2) / (\alpha + 2)$, where $w_\BCKG$ represent the
pressure to energy density ratio of the other background matter fluids
(photons, neutrinos, baryons, and cold dark matter). Moreover, since
we have in this regime $w_\QUINT < w_\BCKG$, this ensures that the
quintessence energy density decreases more slowly than that of the
background fluids and that ultimately, the quintessence field will
become dominant. When this occurs, i.e., when its density parameter
reaches $\Omega_\QUINT \gtrsim 0.5$, the field slows down in its
potential, and reaches asymptotically $w_\QUINT = - 1$ (Steinhardt
\ETAL{} \cite{steinh}). The rate at which one goes from the tracking
regime to the cosmological regime is usually quite slow, so that
unless $w_\QUINT$ is already close to $- 1$ in the tracking regime, it
will still be significantly different from $- 1$ when $\Omega_\QUINT
\sim 0.7$. For example, if $\alpha = 6$, in which case $w_\QUINT = -
0.25$ in the tracking regime during the matter-dominated era, one has
$w_\QUINT = -0.4$ today if $\Omega_\QUINT = 0.7$. This feature leaves
the hope of distinguishing a quintessence field from a cosmological
constant.

The effect of a quintessence field on CMB anisotropies is twofold
(Brax \ETAL{} \cite{brax2}). First, when the quintessence field
becomes dominant it modifies the expansion rate of the universe. This
translates into a modification of the angular distance vs.\ redshift
relation, and hence a shift in the peak structure of the CMB
anisotropies power spectrum, the $C_\ell$'s, for $\ell \gtrsim
100$. Second, the gravitational potentials decay at late time as the
universe is no longer matter-dominated. This produces a so-called
integrated Sachs-Wolfe effect and modifies the $C_\ell$ spectrum at
low multipoles ($\ell \lesssim 20$) as a consequence of the fact that
photons exchange energy with time-varying gravitational potentials.
Both of these effects are also present with a cosmological constant,
but they differ quantitatively with a quintessence field: the shift in
the peak position is smaller, whereas the integrated Sachs-Wolfe
effect can be very different (Caldwell \ETAL{} \cite{caldwell}).

Most of the CMB experiments do not cover a large fraction of the
sky. On the contrary, the new Archeops data (Beno\^\i{}t \ETAL{}
\cite{benoita}; Beno\^\i{}t \ETAL{} \cite{benoitb}) are extremely
precise around $\ell \sim 200$ improving by a factor of two the
precision measurements on the location of the first Doppler peak
(Beno\^\i{}t \ETAL{} \cite{benoitc}), whereas at larger angular scales
(low $\ell$) the COBE data are limited by a large cosmic variance
(Tegmark \cite{tegmark}).  Therefore we can hope being able to
contrain the quintessence parameters by through their influence on the
position of the first Doppler peak rather than through the integrated
Sachs-Wolfe effect. It is well-known that the position of the first
peak is primarily sensitive to the curvature but also to several other
cosmological parameters.  It is therefore important to investigate a
wide space of parameters in order to constraint the possible existence
of quintessence in a robust way.

\begin{figure}[!t]
\begin{center}
\resizebox{\hsize}{!}{\includegraphics[angle=0,totalheight=10cm,
        width=10.cm]{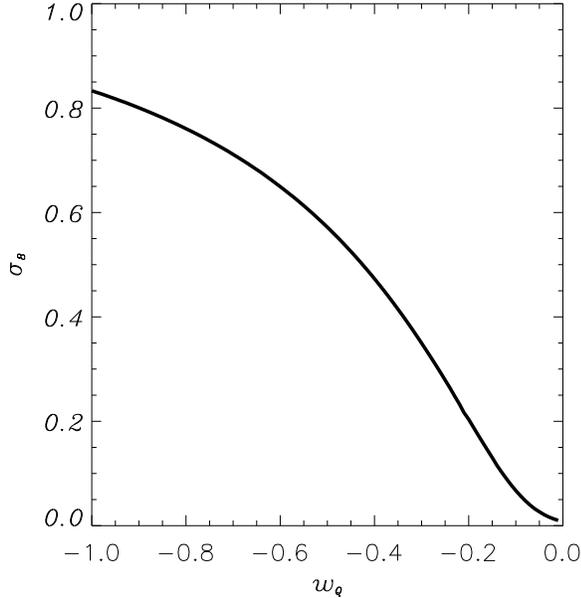}}
\end{center}
\caption{\label{S_W} The amplitude of matter fluctuations $\sig$ as a
function of a constant $w_\QUINT$ in cosmological models with
$\Omega_\QUINT = 0.7$, $\Omega_\BAR h^2 = 0.02$, $H_0 = 66.5$, $n =
1$, and $\Omega_\TOT = 1$, with amplitude normalized to match the COBE
data.  }
\label{fig1}
\end{figure}

Finally, let us add that quintessence also modifies significantly the
matter power spectrum: as for the cosmological constant, matter
fluctuations stop growing at the onset of quintessence domination.
This has an influence on the normalization of the matter power
spectrum on small scales, $\sig$.  This effect of $w_\QUINT$ can be
understood as follows: as long as $w_\QUINT$ is not too close to $0$,
one can roughly consider that the universe has experienced two
distinct epochs since recombination: a first (usual) one where it was
matter-dominated, and a second one where it is
quintessence-dominated. The transition occurs when $\Omega_\MAT =
\Omega_\QUINT = 0.5$. If we suppose that $\Omega_\QUINT = 0.7$ today,
then the transition epoch occured at a redshift of
\begin{equation} 
\nonumber
z^\MAT_\QUINT
 = \left( \frac{\Omega_\QUINT}{\Omega_\MAT} 
   \right)^{\frac{- 1}{3 w_\QUINT}} - 1 .  
\end{equation} 
For $w_\QUINT = -1$ (cosmological constant case), one has
$z^\MAT_\Lambda = 0.33$, whereas for $w_\QUINT = -0.2$, this gives
$z^\MAT_\QUINT = 3.1$. If we suppose that perturbations grow as the
scale factor before the transition and stop growing immediately
afterward, then, this corresponds to a reduction of
\begin{equation} 
\nonumber
R = \left( \frac{1 + z^\MAT_\Lambda}{1 + z^\MAT_\QUINT} \right)^2
  = \left( \frac{\Omega_\QUINT}{\Omega_\MAT} 
   \right)^{\frac{2}{3} \left(1 + \frac{1}{w_\QUINT} \right) } ,
\end{equation} 
in the matter power spectrum as compared to the cosmological constant
case.  In practice, several factors modify this amount, such as the
fact the the matter perturbation still grow (although slowly) in the
quintessence-dominated era, but this illustrate the dramatic change in
the matter power spectrum that quintessence can trigger [see also
Benabed \& Bernardeau (\cite{benber})].  When $w_\QUINT = 0$, the
situation is even more dramatic as matter never dominates. In order to
understand the value of $\sig$, one is has to compute the growth of
dark matter perturbations when one has a mixture of matter and
quintessence with identical equation of state. Neglecting the
quintessence fluctuations (which are never subject to any
instability), the Jeans equation is rewritten as
\begin{equation}
\nonumber
\ddot \delta + {\cal H} \dot \delta - \frac{3}{2} {\cal H}^2
\Omega_\MAT \delta = 0 ,
\end{equation}
When $\delta$ represents the matter perturbation density constrast, an
overdot denotes a derivative with respect to conformal time (defined
by $a \, \ddd \eta = \ddd t$, where $a$ is the scale factor and $t$ is
the cosmic time), and ${\cal H} \equiv \dot a / a$. The only
difference with the usual matter-dominated case is that we do not have
$\Omega_\MAT = 1$ in the above equation. In the matter-dominated era,
$a \propto \eta^2$, so that ${\cal H} = 2 / \eta$, and one has
\begin{equation} 
\nonumber
\ddot \delta + \frac{2}{\eta} \dot \delta - \frac{6}{\eta^2}
\Omega_\MAT \delta = 0 .  
\end{equation} 
When $\Omega_\MAT = 1$, one gets the usual result that the growing
mode goes as $\delta^+ \propto a$. If we now suppose that $\Omega_\MAT
= 0.3$, then the growing mode goes as $\delta^+ \propto a^{0.466}$. If
this regime lasts for several orders of magnitude in redshift, then
this represents an enormous reduction in the matter power spectrum, of
around $(1 + \Delta z)^{1.07}$, where $\Delta z$ represent the
duration of this regime. This effect will imply that $\sig$ is
drastically reduced when one goes from $w_\QUINT = -1$ to $w_\QUINT =
0$.  In order to illustrate this point we have computed the amplitude
of matter fluctuations $\sig$ for a family of models which have the
same same cosmological parameters but the equation of state $w_\QUINT$
(Fig.~\ref{fig1}). As one can see $\sig$ varies dramatically with
$w_\QUINT$ and is therefore a very useful quantity to take into
account in order to constrain quintessence.

\section{Methods and Data}

In the following, we make use of the most recent data available on the
CMB as well as on other relevant cosmological quantities in order to
examine constraints that can be set on the amount of quintessence
present in the universe. We assume Gaussian adiabatic fluctuations and
a flat geometry of space. We assume a vanishing amount of
gravitational waves as such a contribution has little impact on the
position of the peak, but rather modifies the relative amplitude power
between low and large $\ell$. Identically, a reionisation effect and a
possible hot dark matter component are neglected in the following.  We
make the assumption that $w_\QUINT = \CONST$ throughout all the epochs
of interest.  This assumption is unjustified in realistic quintessence
models as one expects $w_\QUINT$ to have varied at the radiation to
matter transition and to be decreasing today. However, if we want to
study the influence of the quintessence field on the position of the
Doppler peaks of the CMB anisotropies, then this is sufficent, and the
constant $w_\QUINT$ has to be seen as some ``average'' of a dynamical
$w_\QUINT (z)$ arising in a quintessence scenario.

\subsection{Likelihood from CMB} 

In order to use CMB data, we first reconstruct the likelihood function
of the various experiments.

We follow the technique developped in Bartlett \ETAL{} (\cite{bdbl}),
and used in Douspis \ETAL{} (\cite{dbsbl}) and Beno\^\i{}t \ETAL{}
(\cite{benoitc}), by constructing a large $C_\ell$ power spectra
database. We investigate six cosmological parameters assuming flat
cosmology ($\OmT = 1$). The density of the universe is parametrised by
the baryon contribution, $\Obhh$, and the dark energy, $\OQ$, for
which we allow its pressure-energy density ratio, $w_\QUINT$ to vary.
 
The Hubble parameter, $H_0$, the spectral index, $n$, and the
normalisation of the spectra, parametrized in this work by $\sig$ are
the remaining free parameters. Table~\ref{grid_table} describes the
corresponding gridding used for the database.

We proceed by estimating cosmological parameters from the likelihood
functions reconstructed as described in Beno\^\i{}t \ETAL{}
(\cite{benoitc}).  We compute the value of the likelihood considering
the actual band powers dataset of the COBE, BOOMERanG, DASI, MAXIMA,
VSA, CBI, Archeops experiments (Tegmark \cite{tegmark}; Netterfield
\ETAL{} \cite{boom2}; Halverson \ETAL{} \cite{halverson}; Lee \ETAL{}
\cite{maxima2}; Scott \ETAL{} \cite{scott}; Pearson \ETAL
\cite{pearson}; Beno\^\i{}t \ETAL{} \cite{benoitb}) on each model of
our grid. In our approach, the best model is estimated as being those
for which the likelihood is maximal ${\cal L}_\MAX$, while the $68\%$
(resp.\ $95\%$) 1-parameter interval corresponds to $-2 \ln({\cal L} /
{\cal L}_\MAX) \in [0, 1]$ (resp.\ $-2 \ln({\cal L} / {\cal L}_\MAX)
\in [0, 4]$) and the $68\%$ (resp.\ $95\%$) 2-parameters interval
corresponds to $-2 \ln({\cal L} / {\cal L}_\MAX) \in [0, 2.3]$ (resp.\
$-2 \ln({\cal L} / {\cal L}_\MAX) \in [0, 6.18]$). The likelihood
shown in the following are already margilalized (by maximisation) over
the calibration uncertainties and the amplitude.  The results are
presented as 2-D contour plots, showing in shades of blue the regions
where the likelihood function for a combination of any two parameters
drops to the levels corresponding to $68\%$, $95\%$, and $99\%$
confidence regions. They would correspond to $1$, $2$, $3$ $\sigma$,
respectively if the likelihood function were Gaussian.  Dashed red
contours mark the limits to be projected if confidence intervals are
sought for any one of the parameters.  To calculate either 1- or 2-D
confidence intervals, the likelihood function is maximized over the
remaining parameters.
\begin{table}[h]
\begin{center}
\begin{tabular}{|c|c|c|c|c|c|c|}
\hline
 & $w_\QUINT$&$\OQ$&$100\;\Obhh $&$H_0$&$n$&$\sig$\\
\hline
\hline
Min. & -1.0   & 0.0  & 0.915  & 25    & 0.750 &  0.1  \\
\hline
Max. & -0.1   & 1.0  & 3.47  & 101   & 1.25 &  1.3 \\
\hline
Step & 0.1   & 0.1  & 0.366  & $\times$1.15 & 0.015 & 0.022 \\
\hline
\hline
\end{tabular}
\end{center}  
\caption{Grid of explored cosmological parameters; for $H_0$ we adopt
a log-periodic binning, $H_0(i+1) = H_0(i) \times 1.15$}
\label{grid_table}
\end{table}

\subsection{Other Data}

In order to combine with other data of cosmological relevance, the
corresponding additional likelihood have to be evaluated.

An interesting useful additional constraint to add is those obtain on
the amplitude of matter fluctuations on small scales: present day
clusters data allow to constrain $\sigma_c$, related to $\sig$ with a
high accuracy, of the order of $5\%$ (Blanchard \ETAL{}
\cite{bsbl}). Similar constraint can be obtained from weak lensing
measurements (Bacon \ETAL{} \cite{bacon}; van Waerbeke \ETAL{}
\cite{vw}). However, significant differences among similar analyses
have appeared in recent works based on clusters as well as on weak
lensing measurements (Jarvis \ETAL{} \cite{jarvis}; Brown \ETAL{}
\cite{brown}, Hamana \ETAL{} \cite{hamana}). We have therefore chosen
to use two recent constraints, the differences of which will allow us
to investigate a realistic range of systematic uncertainties. We first
consider a constraint leading to high values of $\sig$ (high $\sig$
hereafter), in agreement with Pierpaoli \ETAL{} (\cite{pierpaoli}):
$\sig \OmM^{0.6} = 0.5 \pm 10\%$ ($68\%$ C.L.). Then, new estimations
from Seljak \ETAL{} (\cite{seljak}), Viana \ETAL{} (\cite{viana}),
Reiprich \ETAL{} (\cite{reip}), leading to lower values, are
considered by taking $\sig \OmM^{0.38} = 0.43 \pm 10\%$ ($68\%$ C.L.),
hereafter low $\sig$. The latter error bars estimates encompass the
three above low normalization measurements of $\sig$.

We also use the recent determination of the Hubble parameter from the
Hubble Space Telescope (HST) Key Project (Freedman \ETAL{}
\cite{freedman}): $H_0 = 72 \pm 8$ ($68\%$ C.L.), assuming Gaussian
uncertainty.

For the distant supernovae constraints we compute the likelihood as
follows.  The magnitude-redshift relation of the supernovae is given
by the following relation
\begin{equation}
\nonumber
M
 =   {\cal M}
   + 5 \ \log_{10} \ {\cal D}_L(z, w_\QUINT, \Omega_\QUINT) ,
\end{equation}
where $M$ is the observed magnitude and ${\cal D}_L$ and ${\cal M}$
are respectively the ``Hubble-parameter-free'' luminosity-distance and
the ``Hubble-parameter-free'' absolute magnitude at the maximum of the
supernova.  ${\cal D}_L$ is a function of redshift of the supernova,
$z$, and of the cosmological parameters $w_\QUINT$ and $\OQ$.  In the
case of a flat universe with $w_\QUINT$ as the pressure to energy
density ratio of the quintessence field, the ${\cal D}_L$ function is
given by
\begin{eqnarray}
{\cal D}_L (z, w_\QUINT, \OQ)
 & & = \nonumber \\ 
 (1 + z) \int_0^z 
 & &    \frac{\ddd u}
             {\sqrt{  (1 - \OQ) (1 + u)^3
                    + \OQ (1 + u)^{3 (1 + w_\QUINT)}}} .
\end{eqnarray}
With the sample of supernovae from Perlmutter \ETAL{}
(\cite{perlmutter}), we construct the likelihood ${\cal L}_\SN$ with
the following procedure:
\begin{equation}
\nonumber
{\cal L}_\SN
 = C \exp \left[
       \sum_i \left(  - \frac{(  M^\OBS_i
                               - M_i({\cal M}, w_\QUINT, \Omega_\QUINT))^2}
                             {2 \sigma^2_i} \right) \right] ,
\end{equation}
where $C$ is an arbitrary constant.  We take for measured magnitude
$M^\OBS$ and uncertainty $\sigma$ the B-band peak magnitude and the
total uncertainty of the supernovae from Perlmutter \ETAL{}
(\cite{perlmutter}).  At each point of the two parameters grid,
$w_\QUINT$ and $\OQ$ is associated a likelihood value obtained by
maximising it over the absolute magnitude ${\cal M}$.

\section{Results}

\subsection{CMB alone}

\begin{figure}[!t]
\begin{center}
\resizebox{\hsize}{!}{\includegraphics[angle=0,totalheight=10cm,
        width=10.cm]{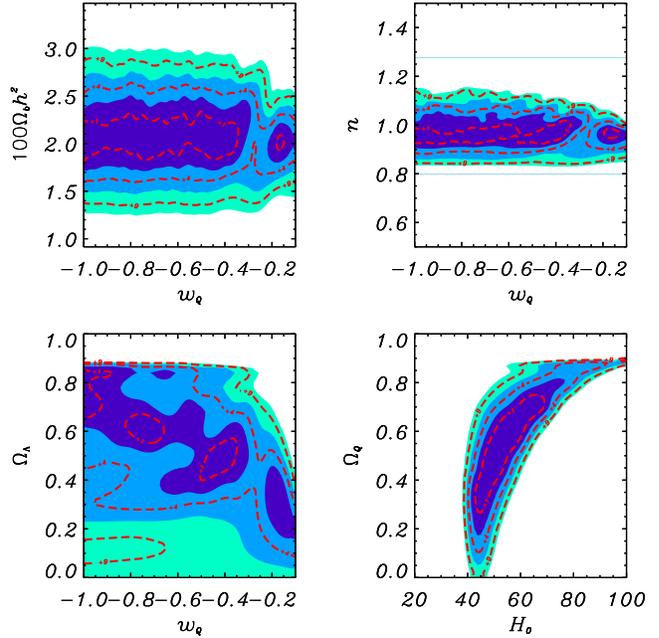}}
\end{center}
\caption{Present CMB dataset likelihood contours in the quintessence
paradigm. The sharpness of contours at $\OQ = 0.9$ is due to grid
effect.}
\label{cmbalone}
\end{figure}    

\begin{figure*}[!t]
\begin{center}
\resizebox{!}{!}{\includegraphics[angle=0,totalheight=5cm,
        width=18.5cm]{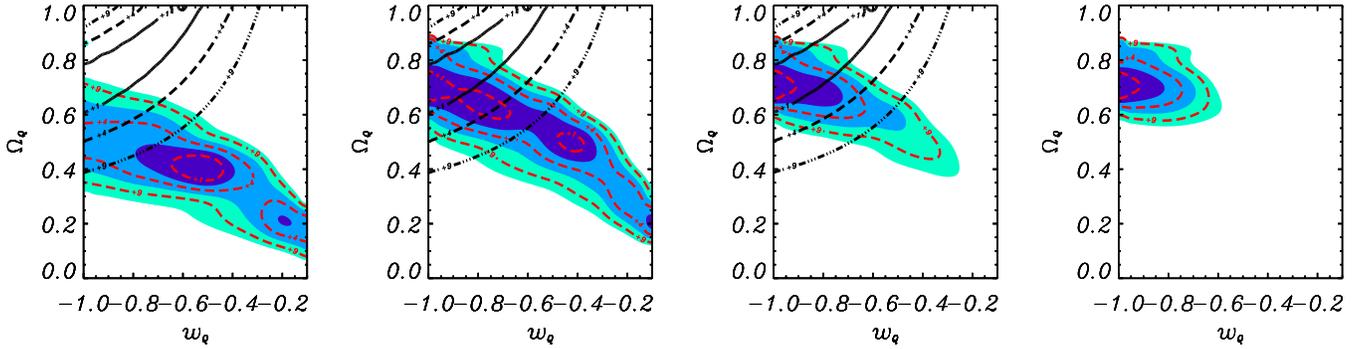}}
\end{center}
\caption{Likelihood contours with combination of CMB and priors. From
left to right, panels show (i) colored contours corresponding to the
combination CMB + $\sig$ (high), wiht overplotted SN contours; (ii)
colored contours corresponding to the combination CMB + $\sig$ (low),
with overplotted SN contours; (iii) colored contours corresponding to
the combination CMB + $\sig$ (low) + HST, with overplotted SN
contours; (iv) colored contours corresponding to the combination CMB +
$\sig$ (low) + HST + SN.}
\label{combi}
\end{figure*}  

Constraints given by the CMB on some of our investigated parameters
are shown in Fig.~\ref{cmbalone}. Considering only CMB constraints
leads to degeneracies between parameters. Fig.~\ref{cmbalone} shows
the case of two parameters, $n$ and $\Obhh$, which are not affected by
the assumed equation of state of the dark energy. Their prefered
values and error bars are $n = 0.95 \pm 0.05$ and $\Omega_\BAR h^2 =
0.021 \pm 0.003$ ($68\%$ C.L.).  Using CMB alone leaves the
2-parameters space ($\OQ, w_\QUINT$) almost unconstrained. Finally, 2D
diagrams $\OQ$ vs.\ $H_0$ or $w_\QUINT$ vs.\ $H_0$ show significant
level of correlation, but with degeneracies. This is illustrated by
the plot $\OQ$ vs.\ $H_0$ shown in Fig.~\ref{cmbalone}.  In our
analysis, we found that with the improvement of CMB data obtained by
the addition of Archeops band powers reduces appreciably the contours
of constraints on the quintessence parameters as well as on
cosmological parameters because of the position and the amplitude of
the first acoustic peak are better determined, but still does not
allow to break the degeneracies.
  
\subsection{Adding Non CMB priors}

As cluster abundance observations lead to a strong constraint on the
normalization of the matter power spectrum, and $\sig$ is rather
sensitive to the changing the equation of state, it is natural to
expect that this constraint in combination with constraints from
$C_\ell$ will lead to tight constraint on quintessence scenarios.
Fig.~\ref{combi} (left panels) shows the combination of CMB data with
the two $\sigma_c$ priors described previously. As one can see,
constraints on the amplitude of matter fluctuations on small scales
has the potential to break the degeneracies between $\Omega_\QUINT$
and $w_\QUINT$. Only a band-shaped region of the plane is not
excluded.  Furthermore, combinations with different priors lead to
different likelihood contours, due to the strong effect of the
equation of state on $\sig$ emphasized in Section 2.  More
specifically, we find that the CMB, combined with the high
normalization leads to a prefered region which is marginally
consistent with the constraints given by high redshift supernovae
(overplotted black lines in Fig.~\ref{combi}). The ``concordance
model'' ($\OQ = 0.7$, $w_\QUINT = - 1$) lies on the $99\%$
C.L. contours, and defines the two dataset as inconsistent. The best
model appears then to have $\OQ = 0.4$, $w_\QUINT = - 0.5$, which is
itself outside the $99\%$ confidence region of the supernovae
constraints.  The combination of CMB with low normalization leads on
the other hand to likelihood contours in agreement with those of
supernovae and HST key project determination of $H_0$.

Due to the form of the joint CMB + $\sig$ (high or low) contours, a
combination with the HST constraints is expected to give stronger
constraints on both $\OQ$ and $w_\QUINT$. The corresponding likelihood
contours of Fig.~\ref{combi} (middle right panel) show that
quintessence is not prefered over classical cosmological constant even
if the degeneracy is not totally broken: $\OQ = 0.70^{+0.16}_{-0.12}$,
$w_\QUINT = -1^{+0.4}$ ($95\%$ C.L.).
 
\begin{figure}[!t] 
\begin{center}
\resizebox{\hsize}{!}{\includegraphics[angle=0,totalheight=5cm,
        width=10.cm]{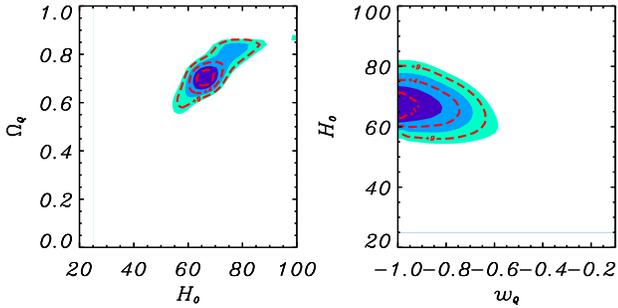}}
\end{center}
\caption{\label{fin}Likelihood contours with CMB + all priors.}
\end{figure}

In order to break the degeneracy, it is clearly necessary to consider
the additional information on the angular distance coming from distant
supernovae. Considering a flat cosmology, the information on the
luminosity of the supernovae can be expressed in term of constraints
on the dark energy density and equation of state. Prefered value are
consistent with a cosmological constant, and the likelihood contours
are almost perpendicular to those of CMB, as shown in
Fig.~\ref{combi}. Combining all the priors finally allows to put
strong constraints on both quintessence parameters (Fig.~\ref{combi},
rightmost panel): $\OQ = 0.70^{+0.10}_{-0.17}$, $w_\QUINT =
-1^{+0.25}$ ($95\%$ C.L.) and finally breaks the $(H_0, \OQ)$
degeneracy, see Fig.~\ref{fin}.

As a main result, it appears that the classical $\Lambda$CDM scenario
is then conforted and given the priors we used there is no need for
quintessence to reproduce the present data, although quintessence
models with low $w_\QUINT$ are still viable, and that good fits to the
data can also be found for models with $w_\QUINT < -1$ (Melchiorri
\ETAL{} \cite{melch})\footnote{During the preparation of this work,
another group submitted a related paper to the archive (Melchiorri
\ETAL{} \cite{melch}). Although the analysis is not exactly the same,
their conclusions are similar.}. However, those correspond to rather
unusual quintessence models.

Going back to typical quintessence models, finding only low values of
$w_\QUINT$ raises some interesting points.  If we consider a pure
inverse power law potential, having $w_\QUINT < -0.6$ when
$\Omega_\QUINT = 0.7$ implies a low value of the exponent $\alpha$ (if
the slope of the potential is too steep, the field does not stop
rapidly when it starts dominating, and $w_\QUINT$ is too large). This
translates into an annoyingly low value of the energy scale $M$
arising in the potential, since one has $M \sim M_\PL (\rho_\CRIT /
M_\PL^4)^{1 / (4 + \alpha)}$ (Brax \ETAL{} \cite{brax2}).  For
example, in order to have $M > 10^3 \UUNIT{TeV}{}$, one needs $\alpha
\geq 6$, whereas data favor $w_\QUINT < -0.6$, which corresponds to
$\alpha \lesssim 3$, or $M \lesssim 20 \UUNIT{GeV}{}$.  There are of
course several possibilities which allow to have a lower $w_\QUINT$ in
quintessence scenarios.  An example to evade this problem is to use
the SUGRA potential found by Brax \& Martin (\cite{brax1}), which is
given by $V = M^{4 + \alpha} / \phi^\alpha \times \exp(\frac{1}{2}
\phi^2 / M_\PL^2)$. The exponential correction comes from supergravity
considerations and insures that the potential has a local minimum
which is almost reached by the field today.  For example, for
$\Omega_\QUINT = 0.7$, one has $w_\QUINT = -0.8$, almost independently
of $\alpha$. This illustrates the fact that if it is still difficult
to distinguish some quintessence models from a cosmological constant
with the present state of available data, although these already allow
to exclude a large number of quintessence models, among which the
simplest one.

\section{Conclusion}

We have studied the constraints that can be obtained on cosmological
parameters within the quintessence paradigm by using various
combinations of observational data set. For simplicity, only models
with constant $w_\QUINT$ were examined as they are believed to be
sufficient to existing data at their present accuracy level. For
similar reasons, we neglected possible reionization or non-zero
gravitational wave contribution.  Indeed, this approach has the
advantage to help to understand whether quintessence models should be
favored or not over a classical cosmological constant, while
constraints on more specific scenarios have to be investigated
directly one by one (Douspis \ETAL{} \cite{dousp2}). Our analysis
method has been to investigate contours in 2-D parameters space. Such
an approach allows to examine possible degeneracies among parameters
which are not easy to identify when constraints are formulated in term
of a single parameter. For instance, we found that CMB data alone,
despite the high precision data obtained by Archeops do not require
the existence of a non-zero contribution of quintessence, because of
the degeneracy with the Hubble parameter: in practice CMB data leave a
large fraction of the $(\Omega_\QUINT, w_\QUINT)$ plane unconstrained,
while only a restricted region of the $(\Omega_\QUINT, H_0)$ plane is
allowed. On the contrary, we found that almost no correlation exist
with the baryonic content $\Omega_\BAR$ nor the primordial index
$n$. In order to restrict the parameter space of allowed models, we
have applied several different constraints. Interestingly, we found
that the amplitude of the dark matter fluctuations, as measured by
clusters abundance or large scale weak lensing data can potentially
help to break existing degeneracies, although existing uncertainties,
mainly systematics in nature, do not allow firm conclusion
yet. Clearly this will be an important check of consistency in the
future. We have then added constraints from supernovae data as well as
HST estimation of the Hubble parameter in order to break existing
degeneracies. This allows us to infer very tight constraints on the
possible range of equation of state of the dark energy. Probably the
most remarkable result is that no preference for quintessence does
emerge from existing CMB data although, accurate measurement of the
amplitude of matter fluctuations on scale of $8 \, h^{-1}
\UUNIT{Mpc}{}$ may change this picture.
 
\begin{acknowledgements} 

The authors acknowledge the use of the CAMB code (Lewis \ETAL{}
\cite{camb}).  M.D.\ is on a CMBNet fellowship and acknowledges Oxford
Astrophysics group computational facilities and the Archeops
collaboration.  Y.Z.\ acknowledges support from the C.N.R.S. This work
has greatly benefitted from discussions inside the Archeops
collaboration

\end{acknowledgements}  

 \end{document}